\documentstyle[twoside,fleqn,espcrc2]{article}

\newcommand{\AmS}{{\protect\the\textfont2
  A\kern-.1667em\lower.5ex\hbox{M}\kern-.125emS}}
\newcommand{\tr}{{\rm Tr}}

\title{Quenched $B_K$-parameter with the Wilson and Clover actions at
$\beta = 6.0$}

\author{M.~Crisafulli$^{\rm a}$,
A.~Donini\address{I.N.F.N. and Dipartimento di Fisica dell' Universit\'a
di Roma ``La Sapienza'' \\ Piazza Aldo Moro 2, I-00185 Roma, Italy}
\thanks{Talk presented by A. Donini}, V.~Lubicz\address{Dept. of Physics,
Boston University, Boston MA 02215,USA.},
G.~Martinelli$^{\rm a}$\address{Theory Division, CERN, 1211 Geneva 23,
Switzerland.}, F.~Rapuano$^{\rm a}$, C.~Ungarelli\address{I.N.F.N. and
Dipartimento di Fisica dell' Universit\'a di Pisa \\ Piazza Torricelli 2,
I-56100 Pisa, Italy}, A.~Vladikas\address{I.N.F.N. and Dipartimento di Fisica
dell' Universit\'a di Roma ``Tor Vergata'' \\ Via della Ricerca Scientifica 1,
I-00133 Rome, Italy}\\
The APE Collaboration\\
A.~Bartoloni$^{\rm a}$, C.~Battista$^{\rm a}$, S.~Cabasino$^{\rm a}$,
N.~Cabibbo$^{\rm e}$, F.~Marzano$^{\rm a}$, E.~Panizzi$^{\rm a}$,
P.S.~Paolucci$^{\rm a}$, R.~Sarno$^{\rm a}$, G.M.~Todesco$^{\rm a}$,
M.~Torelli$^{\rm a}$, P.~Vicini$^{\rm a}$.}
\begin{document}

\begin{abstract}
We present results for the kaon $B$ parameter obtained by APE-100 from a sample
of $100$ configurations using the Wilson action and $200$ configurations
using the Clover action, on a $18^3 \times 64$ lattice at $\beta=6.0$.
We have also compared the results for $B_K$ using two different values for the
effective coupling constant $g$.
We observe a strong dependence of $B_K$ on the prescription adopted for $g$
in the Wilson case, contrary to the results of the Clover case which are
almost unaffected by the choice of $g$.
\end{abstract}

\maketitle

\section{Simulation.}

We have computed the light fermion propagators on an ensemble of
$100$ configurations with the Wilson action and $200$
configurations with the Clover action, on a $18^3 \times 64$ lattice
at $\beta = 6.0$, using the 6-gigaflop version of APE-100.
The light quark masses correspond to the hopping parameter
$K=0.1530,0.1540,0.1550$ for the Wilson action and $K=0.1425,0.1432,0.1440$
for the Clover case. With these values, the meson masses are in the range
$600-900$ MeV. We have considered only pseudoscalar and vector mesons
consisting of quarks degenerate in mass.
Moreover, the two- and three-point functions have been evaluated for mesons
with several momenta. In the three-point functions this corresponds to six
different values of the product $ (p \cdot q) $, $(p,q)$ being the momenta of
the two mesons. All the propagators are thinned. All the errors are jacknife.

%___________________________________________________________________________
\begin{figure}[t]   % produce figure here
    \begin{center}
       \setlength{\unitlength}{1truecm}
       \begin{picture}(6.0,6.0)
          \put(-2.0,-3.5){\special{latt94_bpar_fig1.ps}}
       \end{picture}
    \end{center}
\caption{Dependence of $\langle {\bar K}^0 \mid {\hat O}_{\Delta s = 2}
\mid K^0 \rangle$ on mass and momentum. The points refer to the Wilson action
data.}
\label{massmom}
\end{figure}
%___________________________________________________________________________
%___________________________________________________________________________
\begin{figure}[t]   % produce figure here
    \begin{center}
       \setlength{\unitlength}{1truecm}
       \begin{picture}(6.0,6.0)
          \put(-2.0,-3.5){\special{latt94_bpar_fig2.ps}}
       \end{picture}
    \end{center}
\caption{Comparison between Clover (squares) and Wilson (crosses)
data for mesons at rest.}
\label{comparison}
\end{figure}
%___________________________________________________________________________

\section{The $B_K$ parameter.}

The kaon $B$-parameter is a measure of the deviation from the
vacuum saturation approximation of the matrix element:
\begin{equation}
\label{kok}
\langle {\bar K}^0 \mid {\hat O}^{\Delta S = 2} \mid K^0 \rangle =
\frac{8}{3} f^2_K m^2_K B_K
\end{equation}
where ${\hat O}^{\Delta S = 2} = {\bar s} \gamma^L_{\mu} d {\bar s}
\gamma^L_{\mu} d $.
On the lattice the matrix element (\ref{kok}) is usually parameterized  as
\begin{eqnarray}
\label{koklatt}
\langle {\bar K}^0 \mid {\hat O}_{latt}^{\Delta S = 2} \mid K^0 \rangle & = &
\alpha + \beta m^2_K + \gamma ( p \cdot q ) + \nonumber \\
+ \delta m^4_K +  \epsilon m^2_K (p \cdot q) &+& \zeta (p \cdot q)^2 +
 \dots
\end{eqnarray}
where $\gamma = \frac{8}{3} f^2_K B_K$. The parameters $\alpha,\beta$ are
lattice artefacts that should vanish in the continuum limit.
Following \cite{GAVELA}, in order to reduce the mass dependence of the
different parameters, we have adopted as fit variables:
\begin{eqnarray*}
m^2_K \rightarrow X & = & \frac{8}{3} \frac{f^2_K}{Z_5 Z^2_A} m^2_K
  \\
(p \cdot q) \rightarrow Y & = & \frac{8}{3} \frac{f^2_K}{Z_5 Z^2_A} (p \cdot q)
\end{eqnarray*}
where $Z_5 = <0|\bar \psi \gamma _5 \psi |K^0>$ and $Z_A$ is the axial
current renormalisation constant.
Up to quadratic terms, with $\alpha, \beta, \dots, \zeta$ appropriately
redefined, (\ref{koklatt}) becomes
\begin{eqnarray}
\label{koklatt2}
\frac{1}{ Z_5 } \langle {\bar K}^0 \mid {\hat O}_{latt}^{\Delta S = 2}\mid K^0
 \rangle & = & \alpha + \beta X + \gamma Y + \nonumber \\
+ \delta X^2 & + & \epsilon X Y + \zeta Y^2
\end{eqnarray}
Fitting in $X$ and $Y$ substantially
reduces higher order terms in $(p \cdot q)$ and $m^2_K$ (see \cite{GAVELA}).
When fitting eq.(\ref{koklatt2}), we obtain values for $\alpha,\beta$
and $\gamma$ which differ by about $20\%$ to those obtained from the linear
fit:
\begin{equation}
\label{koklin}
\frac{1}{Z_5} \langle {\bar K}^0 \mid {\hat O}^{latt}_{\Delta s = 2}
\mid K^0 \rangle = \alpha + \beta X + \gamma Y
\end{equation}
This systematic effect is currently under study. Here we only present
results from eq.(\ref{koklin}).
% -----------------------------------------------------
\begin{table*}[hbt]
% space before first and after last column: 1.5pc
% space between columns: 3.0pc (twice the above)
\setlength{\tabcolsep}{1.5pc}
% -----------------------------------------------------
% adapted from TeX book, p. 241
\newlength{\digitwidth} \settowidth{\digitwidth}{\rm 0}
\catcode`?=\active \def?{\kern\digitwidth}
% -----------------------------------------------------
\caption{Clover(SW) and Wilson(W) results with different choices of
effective coupling (see text); $B_K = \gamma / Z_A^2$.}
\label{tab}
\begin{tabular*}{\textwidth}{@{}l@{\extracolsep{\fill}}rrrr}
\hline
 Choice of $g$  & $\alpha$ & $\beta$ & $\gamma$ & $B_K(a^{-1}=2.06 GeV)$ \\
\hline
Unboosted (SW)        & $-0.08(2) $ & $0.3(2)  $ & $ 0.65(15)$ & $0.58(13)$ \\
Boosting Proc. 1 (SW) & $-0.07(2) $ & $0.3(2)  $ & $ 0.65(15)$ & $0.58(13)$ \\
Unboosted (W)         & $-0.09(1) $ & $0.11(8) $ & $ 0.58(8)$  & $0.77(11)$ \\
Boosting Proc. 1 (W)  & $-0.031(5)$ & $0.04(3) $ & $ 0.22(3)$  & $0.48(6)$ \\
Boosting Proc. 2 (W)  & $-0.064(9)$ & $0.08(6) $ & $ 0.42(5)$  & $0.69(8)$ \\
\hline
%\multicolumn{5}{@{}p{120mm}}{Reprinted from: G.M. Ritcey,
%                             Tailings Management,
%                             Elsevier, Amsterdam, 1989, p. 635.}
\end{tabular*}
\end{table*}
% -----------------------------------------------------
In Fig.(\ref{massmom}) the Wilson results are given for
different mass and momentum values.
The set of points at negative values of Y represents the matrix element
$ \langle 0 \mid {\hat O}^{latt}_{\Delta s = 2} \mid K^0 K^0 \rangle$.
We have not used these points in the measure of $B_K$, due to the
final state interaction \cite{MAIANI}.
Points corresponding to mesons with high momenta (i.e. the three points on the
leftmost side of the figure) receive large contributions from the lattice
artefacts, which are to be understood better.
In Fig.(\ref{comparison}) we present a comparison between Wilson and
Clover data for mesons at rest. Some improvement of the chiral
behaviour of the matrix element is obtained by using the Clover action
instead of the Wilson one.

\section{Results.}

We present our Wilson ($W$) and Clover ($SW$) results where, in the
perturbative calculation of the mixing coefficients used in the
renormalisation of ${\hat O}^{\Delta S = 2}$ \cite{martiw},
we have used either the ``naive'' coupling or
two different ``boosting'' procedures for the coupling constant
 \'a la Lepage-Mackenzie \cite{LEPAGE}:
\begin{eqnarray*}
1. \qquad  g^2_1 = (8 K_c)^4 g^2_0 \simeq 2.49 \ (W) ; \ 1.84 \ (SW)
\\
2. \qquad  g^2_2 = \frac{1}{\langle \tr \Box \rangle} g^2_0
\simeq 1.68 \ (W) ; \ 1.68 \ (SW) \nonumber
\end{eqnarray*}
As can be seen, the effective coupling depends quite significantly
on the prescription in the Wilson case, while the
variation in the Clover $(SW)$ case is small.
Our results at $\beta= 6.0$ are shown in the Table, for both the Clover
and Wilson actions and for different boosting procedures. In the Clover
case, procedures 1. and 2. give essentially the same results.
It appears that the Clover value of $\gamma$ is almost unaffected by the
boosting procedure, while the results in the
Wilson case are quite unstable and strongly depend on the choice of the
effective coupling constant. The one-loop renormalisation group invariant
$B$-parameter is defined as:
\begin{equation}
\label{rgi}
{\hat B}_K = \big [ \alpha_s (\mu= 2 GeV)^{-\frac{6}{33-2 N_f}} \big ]
\frac{\gamma}{Z^2_A}
\end{equation}
In this expression we use a non-perturbative estimate of
$Z^{SW}_A = 1.06$ \cite{ZETA_A}, and the perturbative results
$Z^W_A(g^2_0) = 0.87,
Z^W_A(g^2_1) = 0.68, Z^W_A(g^2_2) = 0.78$. From (\ref{rgi}) we obtain:
\begin{itemize}
\item
Clover action:
\begin{equation}
{\hat B}_K = ( 0.74 \pm 0.17 )
\end{equation}
\item
Wilson action:
\begin{equation}
\left \{ \begin{array}{ll}
{\hat B}_K = ( 0.98 \pm 0.14 ) \qquad \mbox{unboosted}
  \\
{\hat B}_K = ( 0.61 \pm 0.08 ) \qquad \mbox{boosted 1}
  \\
{\hat B}_K = ( 0.88 \pm 0.10 ) \qquad \mbox{boosted 2}
\end{array}
\right .
\end{equation}
\end{itemize}
These values reasonably agree with other lattice results \cite{GAVELA},
\cite{BERNARD} (see also A.~Soni and T.~Bhattacharya at this conference)
and are to be compared to other theoretical results, namely those
from QCD sum-rules (QCD/SR)\cite{PRADES} and from
the $1/N_c$ expansion\cite{BARDEEN}.
\begin{equation}
\left \{ \begin{array}{ll}
{\hat B}_K = ( 0.39 \pm 0.10 ) \qquad \mbox{QCD/SR}
  \\
{\hat B}_K = ( 0.70 \pm 0.07 ) \qquad \mbox{$1/N_c$}
\end{array}
\right .
\end{equation}
\par
In conclusion we stress that the non-linear terms in $m^2_K,
(p \cdot q)$ change $B_K$ by about $20\%$.
Moreover, different boosting procedures drastically change the Wilson results,
leaving almost unaffected the Clover ones.
We need to work on larger lattices and remove thinning in order to reduce
finite size effects. It might be useful to adopt some non perturbative
renormalisation procedure in order to improve the chiral behaviour of the
lattice matrix element \cite{NONPER}.

\section*{Acknowledgements.}

We thank R.~Frezzotti and G.~C.~Rossi for useful discussions.


\begin{thebibliography}{9}
%C.~R.~Allton {\it et al.} ; Rome, prep. 94/1050
\bibitem{GAVELA}
M.~B.~Gavela {\it et al.}, Nucl. Phys. {\bf B} 306 (1988) 677.
\bibitem{MAIANI}
L.~Maiani and M.~Testa, Phys. Lett. {\bf B} 245 (1990) 585.
\bibitem{martiw} G. Martinelli, Phys. Lett. {\bf B} 141 (1984) 395;
C. Bernard, A. Soni and T. Draper, Phys. Rev. {\bf D} 36 (1987) 3224.
\bibitem{LEPAGE}
G.P. Lepage and P.B. Mackenzie, Phys. Rev. {\bf D} 48 (1993) 2250.
\bibitem{ZETA_A}
G. Martinelli {\it et al.},  Phys. Lett. {\bf B}311 (1993) 241.
\bibitem{BERNARD}
C.~Bernard and A.~Soni, Nucl. Phys. {\bf B} (Proc. Suppl.) 17 (1990) 495;
R.~Gupta {\it et al.}, Phys. Rev. {\bf D} 47 (1993) 5113;
S.~R.~Sharpe, Nucl. Phys. {\bf B} (Proc. Suppl.) 34 (1994)403.
\bibitem{PRADES}
J.~Prades {\it et al.}, Z Phys. {\bf C} 51 (1991) 287.
\bibitem{BARDEEN}
W.~A.~Bardeen and A.~J.~Buras, J.-M.~Gerard, Phys. Lett. {\bf B} 193 (1987)
138.
\bibitem{NONPER}
C.~Pittori, {\it these proceedings}; G.~Martinelli {\it et al.},
Rome prep. 94/1022
\end{thebibliography}
\end{document}